\documentclass[aps,prl,twocolumn,showpacs,preprintnumbers,amsmath,amsfonts,superscriptaddress]{revtex4}
\usepackage{bbm,array,amsfonts,amsmath,amssymb,dcolumn,bm,nicefrac}
\usepackage[dvips]{graphicx}
\newcommand{\be}{\begin{equation}}
\newcommand{\ee}{\end{equation}}
\DeclareMathOperator{\Tr}{\,{\rm Tr}\,}
\DeclareMathOperator{\tr}{\Tr}

\DeclareMathOperator{\p}{\partial}

\def\({\left(}
\def\){\right)}
\def\[{\left[}
\def\]{\right]}
\def\lb{\lbrace}
\def\rb{\rbrace}

\def\lan{\langle}
\def\ran{\rangle}

\newcommand{\nf}[2]{\nicefrac{#1}{#2}}                    
\def\hbar{\hslash}

\def\re#1{(\ref{eq:#1})}
\newcommand{\nn}{\nonumber}         
\def\hf{\frac{1}{2}}
    
    \def\d{\delta}
    \def\e{\varepsilon}

 \def\la{\lambda}

    \def\s{\sigma}
    \def\t{\tau}

    \def\G{\Gamma}
    \def\D{\Delta}

    \def\Sig{\Sigma}
    
        \def\Bk {{\bf k}}

            \def\Bx {{\bf x}}

       \def\BSig{{\bf\Sigma}}

\newcommand{\GBCS}{{\mathbf{G}_{_\text{\tiny \hspace{-0.7ex}BCS}}}}
\newcommand{\Gm}{{\mathbf{G}_{_\text{\scriptsize \hspace{-0.7ex}m}}}}
\newcommand{\meanfield}{\D}

\newcommand{\Ns}{N_{\rlap{s}\ }}

\begin{document}
\preprint{ITP-UU-06/17}

\title{BEC--BCS crossover in an optical lattice}

\author{Arnaud~Koetsier}
\email{koetsier@phys.uu.nl}
\author{D.~B.~M.~Dickerscheid}
\author{H.~T.~C.~Stoof}

\affiliation{Institute for Theoretical Physics, Utrecht University,
Leuvenlaan 4, 3584 CE Utrecht, The Netherlands}

\date{\today}

\begin{abstract}
We present the microscopic theory for the BEC--BCS crossover of an atomic
Fermi gas in an optical lattice, showing that the Feshbach resonance
underlying the crossover in principle induces strong multiband effects.
Nevertheless, the BEC--BCS crossover itself can be described by a
single--band model since it occurs at magnetic fields that are relatively
far away from the Feshbach resonance. A criterion is proposed for the
latter, which is obeyed by most known Feshbach resonances in ultracold
atomic gases.
\end{abstract}


\maketitle

\emph{Introduction.} ---
Ever since the compelling work of DeMarco and Jin \cite{DeMarco99}, the
study of degenerate Fermi gases has been at the forefront of ultracold
atomic physics. Much of the impetus behind current research in this field
was provided by the successful experimental investigation in the last two
years of the crossover between the Bose--Einstein condensate (BEC) of
molecules and the Bardeen--Cooper--Schrieffer (BCS) state of
Bose--Einstein condensed Cooper pairs
\cite{JILA05,MIT04,Duke04a,Duke04b,Innsbruck04,ENS04,Rice05}. The pairing
observed in an unequal spin mixture is presently under intense scrutiny
\cite{MITunequal,RICEunequal}. Another exciting direction being explored
lately is the physics of ultracold Fermi gases in an optical lattice
\cite{ETH06}. The latter gives, for instance, the possibility to
experimentally solve the famous positive--$U$ Hubbard model with repulsive
on--site interactions which is believed to embody the physics of
high--temperature superconductors \cite{Anderson87,ZhangRice}.
The negative--$U$  Hubbard model with an attractive on--site interaction
\cite{Micnas} can also be explored. This model is interesting for example
because, at half filling, particle--hole symmetry leads to an exact
$SO(4)$ symmetry. Therefore, the physics at half filling is analogous with
that of the $SO(5)$ theory proposed for high--temperature superconductors
\cite{YangZhang,Zhang97,Zhang98}. Moreover, at low filling fractions, a
BEC--BCS crossover takes place. This is the subject of the present Letter.

\begin{figure}[!ht]
\includegraphics[width=\columnwidth]{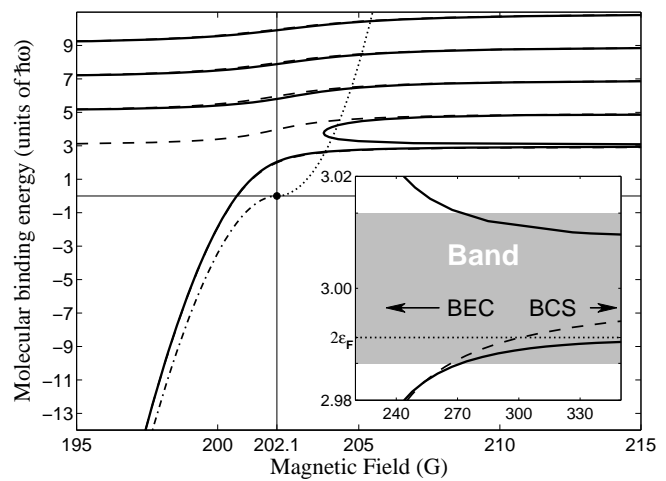}
\caption{Comparison of the dressed molecular binding energy at various
magnetic fields for the homogenous, the on--site two--body, and the
many--body cases corresponding to the dash--dotted, dashed, and solid
lines, respectively. We use ${}^{40}$K atoms near the Feshbach resonance
at $B_0 = 202.1$G and with a filling fraction of 0.1. The on--site
harmonic oscillator frequency $\omega$ is $2\pi\times58275$Hz, which
corresponds to a lattice with wavelength $\la =806$nm and a Rabi frequency
of $\Omega_{\rm R}=2\pi\times1.43$GHz. In the homogeneous case, the
molecules become short lived resonances above $B_0$ and this is indicated
by a dotted curve. The inset shows the two--particle continuum associated
with the lowest Bloch band.
The two--body binding energy crosses through twice the Fermi energy at
$B\simeq 300G$, demonstrating that the BEC--BCS crossover occurs far from
$B_0$. The upper curve in the inset corresponds to the hole--particle
symmetric solution of the BCS gap equation.} \label{mu}
\end{figure}

In the atomic gases of interest, studying the BEC--BCS crossover demands
the exploitation of a Feshbach resonance \cite{Stwalley,Tiesinga}. As a
result, the physics is much richer than in the negative--$U$ Hubbard model
and requires a multiband description. This is elucidated in Fig.~\ref{mu}.
In the absence of an optical lattice, and including solely two--body
effects, a dressed molecule formally exists only below the Feshbach
resonance with a binding energy which, near resonance, depends
quadratically on the magnetic field. Its binding energy is shown as the
dash--dotted line. Above the Feshbach resonance, the dressed molecule is
merely a resonance state as it then has a finite lifetime due to the
possibility of decaying into the atomic continuum \cite{DuinePhD} as
represented by the dotted line continuation.

Placing the system in a sufficiently deep optical lattice, the bare
(closed--channel) molecular state of the Feshbach resonance interacts
primarily with a discrete set of atomic states instead of an atomic
continuum. Thus, a dressed molecule now exists for all magnetic fields and
its energy, plotted as dashed lines, shows a number of avoided crossings
before it ultimately becomes a resonance in the continuum above the
optical lattice potential.

Using this result, we can restate the interacting problem for deep optical
lattices in terms of atoms and dressed molecules emerging from the
on--site two--body solution. The binding energy of these dressed molecules
obtained from our exact solution of the on--site Feshbach problem is in
good agreement with recent experimental data \cite{ETH05,Dicker05-1d},
showing the validity of this approach.
Many--body physics enters if we allow both the atoms and the dressed
molecules to hop to adjacent lattice sites \cite{Dicker05PRA}.
In particular, for the low filling fractions considered, the two--body
binding energy of the dressed molecules passes through the Fermi sea in
the lowest Bloch band, as shown in the inset of Fig.~\ref{mu}. It is due
to this phenomenon that the BEC--BCS crossover occurs in this resonantly
interacting case \cite{FalcoStoof}. The interaction of the molecules with
the Fermi sea further clads the dressed molecules with many--body effects
resulting in a change of the binding energy such that it only
asymptotically reaches twice the Fermi energy at high magnetic fields, as
shown by the solid line. Mathematically, this arises because the two--body
pole at $3\hbar\omega$ in the self--energy of the molecules is replaced by
the famous logarithmic BCS singularity at twice the Fermi level.
By placing the system in an optical lattice, we thus find that the
BEC--BCS crossover takes place rather far from the resonance.
Consequently, as we shall show below, we may always use the single--band
approximation to describe the BEC--BCS crossover for all experimentally
relevant Feshbach resonances.

\bigskip\emph{BEC--BCS crossover theory in the lattice.} ---
We now begin with an outlay of our general BEC--BCS crossover theory in
the lattice before discussing the specific case of ${}^{40}$K. The
second--quantized grand--canonical Hamiltonian in the presence of an
optical lattice potential $V(\Bx)$ is
\begin{align}
H=& \int\!\! d\Bx\; \psi_{\rm m}^\dag(\Bx)
\( - \frac{\hbar^2 \nabla^2}{4m} + \d - 2\mu + 2V(\Bx)\)
\psi_{\rm m}(\Bx)\nn\\
&+\ \sum\limits_{\s=\uparrow,\downarrow} \int\!\! d\Bx\; \psi^\dag_{\s}(\Bx)
\(- \frac{\hbar^2 \nabla^2}{2m} - \mu + V(\Bx)\)
\psi_{\s}(\Bx)\nn\\
&+\ g\int\!\! d\Bx\( \psi_{\rm m}^\dag(\Bx) \psi_{\uparrow}(\Bx) \psi_{\downarrow}(\Bx) +
\psi^\dag_\downarrow(\Bx) \psi^\dag_\uparrow(\Bx) \psi_{\rm m}(\Bx)\), \label{eq:ham}
\end{align}
where $\psi_{\s}(\Bx)$ are the atomic and $\psi_{\rm m}(\Bx)$ the
molecular annihilation operators, and $\mu$ is the chemical potential. The
atom--molecule coupling is given by $g=\hbar\sqrt{4\pi a_{\rm bg}\D
B\D\mu_{\rm mag}/m}$, where $a_{\rm bg}$ is the background scattering
length, $\D B$ is the width of the Feshbach resonance, $\D\mu_{\rm mag}$
is the difference in magnetic moments between the closed and open channels
of the Feshbach resonance, and $m$ is the atomic mass
\cite{DuinePhD,FalcoStoof}. The detuning from resonance is $\d=\D\mu_{\rm
mag}(B-B_0)$, where $B_0$ is the value of the magnetic field $B$ at the
resonance. It gives the location of the bare molecular level with respect
to the atomic continuum in the absence of the lattice.

With a lattice potential present, we turn to a description in terms of
Bloch bands since plane wave states no longer diagonalize the Hamiltonian.
Summing the effects of nearest--neighbour tunneling leads to a dispersion
in the lowest band of the optical lattice equal to $\e(\Bk)=-2t^{\rm a}
[\cos(k_x\la/2)+\cos(k_y\la/2)+\cos(k_z\la/2)]+3\hbar\omega/2$, with $\la$
the wavelength of the lattice laser and $3\hbar\omega/2$ the on--site,
ground--state energy of a single atom. The atomic and molecular tunneling
amplitudes are given by $ t_{\mathrm{a,m}} = 4
    (V_0^3 E^{\mathrm{R}}_{\mathrm{a,m}}/\pi^2)^{\nf{1}{4}}
    \exp(-2\sqrt{V_0/E^{\mathrm{R}}_{\mathrm{a,m}}})
$, where $V_0$ is the peak--to--trough depth of the optical lattice
potential and the atomic and molecular recoil energies are
$E^{\mathrm{R}}_{\mathrm{a}}=2(\pi\hbar)^2/m\la^2$ and
$E^{\mathrm{R}}_{\mathrm{m}} = E^{\mathrm{R}}_{\mathrm{a}}/2$. Moreover,
the effective on--site atom--molecule coupling $g_b^2\ =\ g^2\int
d\Bx|\chi_b(\Bx)|^4$ becomes band--dependent, where $\chi_b(\Bx)$ is the
Wannier function in band $b=0,1,2,\ldots$ of the optical lattice. In the
lowest band, we write this effective coupling as $g'\equiv g_0=g/(2\pi
l^2)^{\nf{3}{4}}$ where $l=\sqrt{\hbar/m\omega}$ is the on--site harmonic
oscillator length \cite{Dicker05PRA}.

Obtaining pertinent thermodynamic quantities for this system involves the
calculation of the molecular self--energy. We obtain this self--energy by
first integrating the fermions out of the partition function exactly. We
then perform an RPA or Bogoliubov approximation around the mean field $\D
= g'\lan\psi_{\rm m}\ran \equiv g'\sqrt{n_{\rm mc}^{\rm B}}$ of the bare
molecular condensate. In the on--site two--body limit, we have $t_{\rm
a}=0$ and $\D=0$, leaving dispersionless harmonic oscillator bands with
energies $\e_{b}=(b+3/2)\hbar\omega$. The two--body molecular self--energy
then becomes $\hbar\Sig(E)=g'^2\sqrt{\pi}G(E-3\hbar\omega)/\hbar\omega$
where the analytic sum over dispersionless bands is encapsulated in the
function \cite{Busch98} $G(E)\equiv \G(-E/2\hbar\omega)/\G(-E/2\hbar\omega
- 1/2)$. The binding energy of the dressed molecules is then found from
$E=\d+3\hbar\omega/2+\hbar\Sig(E)$.

In the many--body case, where $t_{\rm a}$, $t_{\rm m}$, and $\D$ are all
nonzero in general, we work in a regime where the temperature is
sufficiently below the Fermi temperature that we may use a
zero--temperature approximation.
Due to the large band gap in a deep optical lattice, the structure of all
bands but the lowest has a negligible effect and they can be taken to be
dispersionless, as in the two--body case. However, they can not be
excluded altogether from calculations since, as we have just witnessed,
they renormalize the two--body energy levels. The BCS dispersion relation
of the fermions is now $\hbar\omega(\Bk)=\sqrt{(\e(\Bk)-\mu)^2+|\D|^2}$
and we can write the normal and anomalous molecular self--energies in
terms of the usual coherence factors of BCS theory $|u(\Bk)|^2 =
[\hbar\omega(\Bk) + \e(\Bk) - \mu]/2\hbar\omega(\Bk)$ and $|v(\Bk)|^2 =
[\hbar\omega(\Bk) - (\e(\Bk) - \mu)]/2\hbar\omega(\Bk)$ to obtain
\begin{widetext}
\begin{align}
\hbar\BSig_{11}(\Bk,i\omega_n) &= \frac{g'^2 \sqrt{\pi} G(E)}{\hbar\omega}
-\frac{g'^2}{i\hbar\omega_n \!-\! 3\hbar\omega \!+\! 2\mu} + \frac{g'^2
}{\Ns} \!
    \sum_{\Bk'}
\Bigg\lb\!\! \frac{ |u(\Bk_+)|^2 |u(\Bk_-)|^2}{
i\hbar\omega_{n}\!-\!\hbar\omega(\Bk_+)\!-\!\hbar\omega(\Bk_-)} - \frac{
|v(\Bk_+)|^2 |v(\Bk_-)|^2}{
i\hbar\omega_{n}\!+\!\hbar\omega(\Bk_+)\!+\!\hbar\omega(\Bk_-)}\!\!\Bigg\rb,
\label{eq:SE11} \\
\hbar\BSig_{12}(\Bk,i\omega_n) &= \frac{2g'^2}{\Ns}
    \sum_{\Bk'}
u(\Bk_+) v(\Bk_+) u(\Bk_-) v(\Bk_-) \frac{\hbar\omega(\Bk_+) +
\hbar\omega(\Bk_-)}{[\hbar\omega(\Bk_-) + \hbar\omega(\Bk_+)]^2 +
\hbar^2\omega^2_n}, \label{eq:SE12}
\end{align}
\end{widetext}
where $\Bk_\pm=\Bk'\pm\Bk/2$, $E=i\hbar\omega_n+2\mu-3\hbar\omega$,
$N_{\rm S}$ is the total number of lattice sites and the sums are over the
first Brillouin zone only: $k'_x,k'_y,k'_z\in(-2\pi/\la,2\pi/\la)$. An
equation for the BCS gap $\D$ is then found in the form of a
Hugenholtz--Pines relation
\be
2\mu=\d + \frac{3\hbar\omega }{2} - 6t^{\rm m} +
\hbar\BSig_{11}(\mathbf{0},0)-\hbar\BSig_{12}(\mathbf{0},0).
\label{eq:gap}
\ee

Having these quantities in hand enables us to determine the partition
function of the system from the Gaussian effective action of our RPA
theory and, via the resulting thermodynamic potential $\Omega$, we can
calculate the total number of particles using the identity
$N=-{\p\Omega}/{\p\mu}$. This leads to the equation of state for the total
atomic filling fraction $n=N/N_{\rm S}$ given by
\be\label{eq:eos}
n= \tr\[\t_3\GBCS \] + 2 \frac{|\meanfield|^2}{g'^2} - \tr\[ \Gm \] +
\hf\tr\[ \Gm\frac{\p\hbar\BSig}{\p\mu} \],
\ee
where $\t_3$ is the third Pauli matrix and the $2\times2$ Nambu space
matrices $\GBCS$, $\Gm$ and $\BSig$ are the BCS atomic Greens function,
the molecular Greens function, and self--energy matrix composed of
Eqs.~\re{SE11} and \re{SE12} respectively.

\begin{figure}[!ht]
\includegraphics[width=0.9\columnwidth]{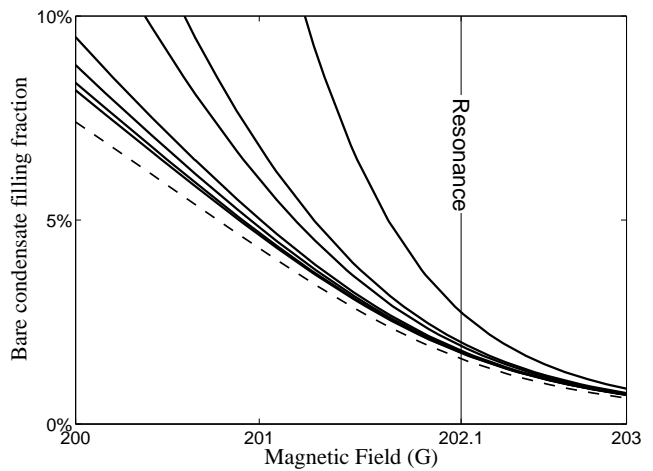}
\caption{The filling fraction of the bare molecular condensate, $2n_{\rm
mc}^{\rm B}/n$, drawn as solid lines, decreases monotonically to the
two--body limit as progressively more bands are included, from 1 to 300 in
this case. The dashed line is $Z^{\rm 2B}$, which is equivalent to the
filling fraction of the molecular condensate in the BEC region shown
here.} \label{nmcb}
\end{figure}

\bigskip \emph{Results and discussion.} ---
Although our approach is quite general, the numerical calculations in this
Letter are for the $B_0=202.1$G Feshbach resonance of ${}^{40}$K atoms,
which has a width of $\D B=7.8$G \cite{JILA05}, and for a total filling
fraction of 0.1 as an illustrative and experimentally relevant case. For
every $\d$, solving the BCS gap equation, Eq.~\re{gap}, and the equation
of state simultaneously yields self--consistent values for $\mu$ and the
filling fraction of the bare molecular condensate $n_{\rm mc}^{\rm B} =
\D^2/g'^2$. Since fluctuation effects are expected to be less important
away from resonance, in the first instance we neglect the fluctuation
contribution reflected by the last two terms on the right--hand side of
Eq.~\re{eos}. In order to examine $n_{\rm mc}^{\rm B}$, it proved
requisite to include the structure of the higher bands to which
intermediate states strongly couple in the neighbourhood of the Feshbach
resonance. Having done so, the many--body result monotonically approaches
the 2--body limit in the resonant region as more bands are included in the
self--energy calculation. This is clear from Fig.~\ref{nmcb}. When
including higher bands, we used a multiband generalization of
Eqs.~(\ref{eq:SE11}) and (\ref{eq:SE12}) that did not factor in any
interband coupling other than the already present interaction of the mean
field $\D_b=g_b\lan\psi_{\rm m}\ran$ with the molecular condensate.
Although this leads to only qualitative results in the higher bands, the
correct two--body physics is obtained and the importance of multiband
effects near resonance is well demonstrated.

\begin{figure}[!ht]
\includegraphics[width=\columnwidth]{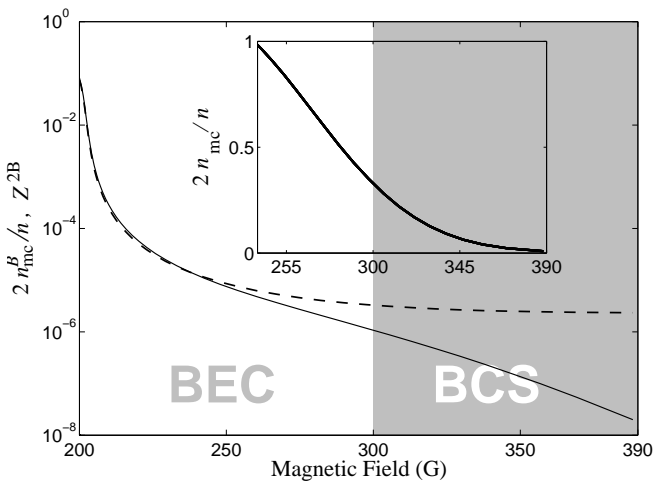}
\caption{Two--body wavefunction renormalisation factor $Z^{\rm 2B}$
(dashed line) plotted together with the filling fraction of the bare
molecular condensate $2n_{\rm mc}^{\rm B}/n$ including a varying number of
bands (solid lines), which shows that a single--band calculation is
adequate in the BEC--BCS crossover region of interest. The inset shows the
approximate filling fraction of dressed molecules, $n_{\rm mc}$, in the
crossover region. The multiple solid lines of the main graph all lie on
top of each other in this high magnetic field region.} \label{z}
\end{figure}

We also determine the filling fraction of Bose--Einstein condensed dressed
molecules as this will provide us with another good indication of where
the crossover from a Bose--Einstein condensate of dressed molecules to the
BCS state composed of Bose--Einstein condensed Cooper pairs takes place.
This fraction is given by $n_{\rm mc}=n_{\rm mc}^{\rm B}/Z$, where $Z$ is
the so--called wavefunction renormalisation factor which determines the
probability amplitude for the dressed molecular wavefunction to be in the
bare molecular state \cite{DuinePhD}. This is plotted in the inset of
Fig.~\ref{z}.
For this purpose, we did not calculate the true many--body $Z$, which
would involve the full fluctuation calculation \cite{Romans05}, but used
instead the 2--body wavefunction renormalisation factor given by $Z^{\rm
2B}=1/(1-\p \hbar\Sig(E) / \p E)$, where $\hbar\Sig(E)$ is the two--body
self--energy discussed earlier.
This results in an upper bound for $n_{\rm mc}$, owing to the fact that
$Z^{\rm 2B}$ is always smaller than its many--body counterpart $Z$. The
magnetic field region where this dressed filling fraction decreases from
one to zero indicates the position of the BEC--BCS crossover, which is
clearly in the same neighbourhood as where the two--body molecular energy
enters the Fermi sea. From the inset in Fig.~\ref{z}, we see that the
BEC--BCS crossover is dominated by contributions from the lowest band
only; including higher bands in the self--energy calculation does not
bring about any apparent change.

In Fig.~\ref{z} we compare $Z^{\rm 2B}$ with $2n_{\rm mc}^{\rm B}/n$.
Indeed, $2n_{\rm mc}^{\rm B}/n$ and $Z^{\rm 2B}$ are similar in the BEC
limit, since the gas then consists solely of a Bose--Einstein condensate
of dressed molecules. This is a result of the fact that in this regime the
energies of the atomic states lie far above the dressed molecular state,
as seen in Fig.~\ref{mu}, which suppresses many--body effects. The BCS
region, however, sees a pronounced difference between the two--body
renormalisation factors and the bare molecular condensate fraction.

Our solutions of the BCS gap equation confirm the presence of the
logarithmic BCS singularity arising at high detuning where the gap then
scales as $\D \simeq 2t^\s\exp(-29\,t^\s/|U_{\rm eff}|)$ \cite{Micnas}.
Here, $U_{\rm eff}\simeq -g'^2/(\d+3\hbar\omega-2\mu)$, is the on--site
Hubbard attraction of the atoms. In the crossover region, the molecular
energy crosses the lowest--band boundary when $\d\simeq g'^2/12t_{\rm a}$.
Furthermore, $U_{\rm eff}$ indicates this crossing occurs close to a
Feshbach resonance when $\d\simeq3\hbar\omega/2$. This leads to the
criterion 
$\D B\gg 18t_{\rm a}\sqrt{4\pi\hbar/m\omega}/a_{\rm bg}\D\mu_{\rm mag}$
which ensures that the Feshbach resonance and the BEC--BCS crossover are
well separated, as required for the applicability of a single--band model
to the BEC--BCS crossover in an optical lattice. Most known Feshbach
resonances adhere to this criterion.

In conclusion, we have presented the full microscopic many--body theory
that describes the BEC--BCS crossover in an optical lattice due to the
presence of a Feshbach resonance. Both two--channel physics and the
possible effects of higher bands were taken into account. The many--body
$Z$ can only calculated from the full fluctuation theory and this is an
important topic of further study.

We would like to thank Mathijs Romans, Michiel Snoek, and Usama Al Kahwaja
for helpful discussions. This work is supported by the Stichting voor
Fundamenteel Onderzoek der Materie (FOM) and the Nederlandse Organisatie
voor Wetenschaplijk Onderzoek (NWO).


\end{document}